\documentclass[fleqn,usenatbib]{mnras}

\usepackage{newtxtext,newtxmath}
\usepackage[T1]{fontenc}
\usepackage{ae,aecompl}
\usepackage{graphicx}	% Including figure files
\usepackage{amsmath}	% Advanced maths commands
\usepackage{amssymb}	% Extra maths symbols
\usepackage[caption=false]{subfig}
\title[GX 339--4 in quiescence]{Radio \& X-ray detections of GX 339--4 in quiescence using MeerKAT and \textit{Swift}} 
\author[E. Tremou et al.]{E. Tremou,$^{1}$\thanks{E-mail: evangelia.tremou@cea.fr}
S. Corbel,$^{1,2}$
R.P. Fender,$^{3,4}$
P.A. Woudt,$^{4}$
J.C.A. Miller-Jones,$^{5}$ 
\newauthor 
S.E. Motta,$^{3}$
I. Heywood,$^{3,6}$
R. P. Armstrong,$^{3,4,7}$
P. Groot,$^{4,8,9}$
A. Horesh,$^{10}$
\newauthor 
A. J. van der Horst,$^{11,12}$
E. Koerding,$^{9}$
K. P. Mooley,$^{13,14,15}$
A. Rowlinson,$^{16,17}$
\newauthor
and R. A. M. J. Wijers$^{16}$
\\ \\
$^{1}$AIM/CEA Paris-Saclay, Universit\'e Paris Diderot, CNRS, F-91191 Gif-sur-Yvette, France\\
$^{2}$Station de Radioastronomie de Nan\c cay, Observatoire de Paris, PSL Research University, CNRS, Univ. Orl\'eans, 18330 Nan\c cay, France\\
$^{3}$Astrophysics, Department of Physics, University of Oxford, Keble Road, Oxford OX1 3RH, UK\\
$^{4}$Inter-University Institute for Data-Intensive Astronomy, Department of Astronomy, University of Cape Town, Private Bag X3, \\ Rondebosch  7701, South Africa\\
$^{5}$International Centre for Radio Astronomy Research, Curtin University, GPO Box U1987, Perth, WA 6845, Australia\\
$^{6}$Department of Physics and Electronics, Rhodes University, PO Box 94, Grahamstown 6140, South Africa\\
$^{7}$South African Radio Astronomy Observatory, 2 Fir Street, Black River Park, Observatory, Cape Town 7925, South Africa\\
$^{8}$South African Astronomical Observatory, PO Box 9, Observatory 7935, South Africa\\
$^{9}$Department of Astrophysics/IMAPP, Radboud University Nijmegen, P.O. Box 9010, 6500 GL Nijmegen, The Netherlands\\
$^{10}$Racah Institute of Physics, The Hebrew University of Jerusalem, Jerusalem 91904, Israel\\
$^{11}$Department of Physics, The George Washington University, 725 21st Street NW, Washington, DC 20052, USA\\
$^{12}$Astronomy, Physics and Statistics Institute of Sciences (APSIS), 725 21st Street NW, Washington, DC 20052, USA\\
$^{13}$Department of Physics, University of Oxford, Keble Road, Oxford OX1 3RH, UK\\
$^{14}$National Radio Astronomy Observatory, Socorro, NM 87801, USA\\
$^{15}$Caltech, 1200 E. California Blvd. MC 249-17, Pasadena, CA 91125, USA\\
$^{16}$Anton Pannekoek Institute, University of Amsterdam, Postbus 94249, 1090 GE, Amsterdam, The Netherlands\\
$^{17}$Netherlands Institute for Radio Astronomy (ASTRON), Oude Hoogeveensedijk 4, 7991 PD, Dwingeloo, The Netherlands\\
}

\date{Accepted 2020 February 3. Received 2020 January 17; in original form 2019 December 5}

\pubyear{2020}

\begin{document}
\label{firstpage}
\pagerange{\pageref{firstpage}--\pageref{lastpage}}
\maketitle

\begin{abstract}
The radio:X-ray correlation that characterizes accreting black holes at all mass scales - from stellar mass black holes in binary systems to super-massive black holes powering Active Galactic Nuclei - is one of the
most important pieces of observational evidence supporting the existence of a connection between the accretion process and the generation of collimated outflows - or jets - in accreting systems.
%Determining the low luminosity fraction of the radio:X-ray luminosity plane in the X-ray binary systems during their quiescence state has substantial implications to their physical interpretation. The correlation between radio and X-ray luminosity is a powerful tool to investigate the connection between accretion and the generation of relativistic jets in these systems. 
Although recent studies suggest that the correlation extends down to low luminosities, only a handful of stellar mass black holes have been clearly detected, and in general only upper limits (especially at radio wavelengths) can be obtained during quiescence.
We recently obtained  detections of the black hole X-ray binary GX 339--4 in quiescence using the MeerKAT radio telescope and \textit{Swift} X-ray Telescope instrument on board the Neil Gehrels Swift Observatory, probing the lower end of the radio:X-ray correlation. We present the properties of accretion and of the connected generation of jets in the poorly studied low-accretion rate regime for this canonical black hole XRB system. 
\end{abstract}


%%%%%%%%%%%%%%%%%%%%%%%%%%%%%%%%%%%%%%%%%%%%%%%%%%

%%%%%%%%%%%%%%%%% BODY OF PAPER %%%%%%%%%%%%%%%%%%

\section{Introduction}

X-ray binaries (XRBs) are binary systems comprised of a compact stellar remnant (a black hole or a neutron star) and a companion star with active mass accretion onto the stellar remnant. The presence of the collapsed star is revealed by X-ray and radio activity whose (relative and absolute) strength depends on the accretion rate onto the compact object and the state of the accretion disk that forms around the compact object. In low-mass XRBs, the accretion from a low mass donor star occurs through Roche-lobe overflow: matter streams from the companion star to the compact one, forming an accretion disc that redistributes angular momentum and emits copious radiation peaking in the X-rays.
%in contrast with high mass X-ray binaries, where the compact star is fed through stellar wind capturing. 

Transient XRBs spend most of their lives in a so-called \textit{quiescent state}, during which they accrete at very low mass-accretion rates and emit at X-ray luminosities ranging between 10$^{30}$ and 10$^{33}$ erg s$^{-1}$ \citep{2002kong,2008gallo}. %Powerful jets are known to occur in the so-called \textit{hard states} of black hole XRBs \citep{2001fender}, the accretion state immediately following and preceding the quiescent state. %Such jets appear to dominate the energetics of the accretion flow at low accretion rates \citep{2003fender}, and may even be responsible for part of the X-ray emission \citep{2003markoff,2004markoff} during the hard states.
The quiescent state is interrupted by occasional \textit{outbursts}, active phases that can last from weeks to years, during which the source's X-ray luminosity can reach or even cross the  Eddington limit, L$_{Edd}$ \citep{2000king}. The accretion rate and luminosity of XRBs, and their X-ray spectral and fast time-variability properties, change dramatically as they go through the quiescence/outburst cycle \citep{2006remillard}. 

While an optically thick, geometrically thin accretion disk (thermal emission) dominates the X-ray emission at high accretion rates, non-thermal radiation from a radiatively inefficient accretion flow located in the inner regions of the accretion disk dominates at lower rates \citep{2014yuan}. The radiatively inefficient accretion flow is believed to be linked to the generation of compact jets, which emit synchrotron radiation, with a peak luminosity in the radio regime \citep{2000corbel, 2001fender}. Such jets appear to dominate the energetics of the system at low accretion rates \citep{2003fender}, and may even be responsible for part of the X-ray emission \citep{2003markoff,2004markoff} during the hard states. As a consequence, the radio and X-ray emission at low accretion rates are tightly correlated  \citep[e.g:][]{1998hannikainen,2000corbel,2003corbel,2013corbel, 2003gallo, 2012gallo, 2014gallo, 2011coriat, 2016tetarenko}, and the radio:X-ray correlation is one of the
most important pieces of observational evidence supporting the existence of a correlation between accretion and the generation of jets, also called \textit{disc-jet coupling}. This correlation encompasses accreting compact objects at all scales (thus including super-massive black holes in Active Galactic Nuclei, AGN), when a mass term is considered, and is often referred to as \emph{the fundamental plane of activity of black holes} \citep{2003merloni,2004falcke}. %This existence of a disc-jet coupling implies that an increase in the mass accretion rate onto the compact object during an outburst, which results in an increase in the X-ray emission, may trigger an increase in the mass loaded into the jet, thus resulting in an increase in the radio emission as well. 

While the outburst phases of black hole XRBs are relatively well studied, this is not true for quiescence and the low-luminosity states in general, which are seldom probed due to the difficulties in observing simultaneously these sources at very low fluxes. During such states, one has the opportunity to study the accretion process and the accretion-powered jets in a regime where the thermal emission from the thin disc does not outshine the radiation from non-thermal processes, which might have a fundamental role in the disc-jet coupling. Also, it has been recently shown that the quiescent state can show significant variability \citep[e.g:,][]{2008millerjones,2009hynes,2011froning,2016wu,2019plotkin}.

Interestingly, differences in the quiescent emission of neutron star versus black hole transients have been proposed as a possible signature of the absence of a hard surface, and might thus provide an indirect evidence for the existence of an event horizon \cite[e.g:][although see \cite{2006jonker} for difficulties with this interpretation]{1999quataert,1999menou,2003mcclintocks}. Hence, studying  XRBs at low accretion rates is key to determine what properties - if any - of the disc-jet coupling depend on the nature of the compact object powering these systems. 
%Furthermore, the studies of XRBs at low accretion rates are interesting because the difference in quiescent emission of neutron star versus black hole transients has been suggested as a possible signature of the absence of a hard surface, and thus the presence of an event horizon (e.g. Quataert \& Narayan 1999; Menou et al. 1999; McClintock et al. 2003; although see Jonker et al. 2006 for difficulties with this interpretation).

%Furthermore, powerful jets are known to occur in the so-called \textit{hard states} of black hole XRBs \citep{2001fender}, the accretion state immediately following and preceding the quiescent state. Such jets appear to dominate the energetics of the accretion flow at low accretion rates \citep{2003fender}, and may even be responsible for part of the X-ray emission \citep{2003markoff,2004markoff} during the hard states. %Determining the contribution to the jet to the X-ray emission is therefore essential to estimate the total energy output of jetted, accreting collapsed objects at low accretion rates, a piece of information that extrapolated to the AGN scales could shed light on the feedback processes at play around low-accretion rate AGN. 

GX 339--4 is a Galactic XRB with a low mass companion ($\sim$1M$_{\odot}$) orbiting a central black hole with mass  $\geq$5.8M$_{\odot}$ \citep{2004hynes}, with an orbital period of 42 hours, located at a distance of 8 - 12 kpc \citep{2019zdziarski}. GX 339--4 is one of the best studied black hole X-ray transients, having been observed in the radio and X-rays during several different outbursts over more than 40 years. GX 339--4 is also one of the key sources in the radio:X-ray correlation, as it is currently the system featuring the best simultaneous X-ray and radio data-sets covering several outbursts \citep{2013corbel}. Here we present only the detection in quiescence. 

%In Section \ref{obs}, we present the data of our radio and X-ray observations that we use for this analysis presented in Section \ref{res} together with our results. Finally, we discuss and summarize our conclusions in Section \ref{disc}.

\section{Observations}\label{obs}
As part of the large survey project ThunderKAT \citep{2017fender}, we are studying a large number of radio transients, including many XRBs, in the image-domain .  The field of the XRB GX 339--4 is observed weekly since September 2018 with the full MeerKAT \cite[Meer Karoo Array Telescope;][]{2009jonas} array. A dedicated \textit{Neil Gehrels Swift Observatory} \cite[hereafter referred to as \textit{Swift};][]{2004gehrels} monitoring program supports these observations, providing weekly X-ray measurements. %The source was in quiescence state from April 2018 till the end of November 2018. 

\subsection{MeerKAT radio observations}

 The  observations discussed here were taken between April 2018, when the first ThunderKAT data were obtained using the full MeerKAT array, and November 2018, when the quiescent phase of the source ended with an outburst. The MeerKAT radio telescope is located in the Karoo desert in South Africa and comprises 64 antennas, 13.5 meters diameter each, with a maximum baseline of 8 km. Observations were made using the L-band (900 - 1670 MHz) receiver, split into 4096 frequency channels spanning 856 MHz centered at 1284 MHz.  Observations typically alternated between the target and phase calibrator, while a bandpass and flux calibrator was observed at the beginning of the observing block. All observations were obtained in full polarization mode. The data were calibrated following the standard procedures with the Common Astronomy Software Application \citep[CASA;][]{2007mcmulin}. Imaging, self-calibration and direction-dependent calibration of the data were carried out with the new wide-band, wide-field imager, DDFacet \citep{2018tasse}. For details, see section 2, \cite{2020driessen}.

\begin{table}
\tiny
\caption{MeerKAT \& \textit{Swift} observations of GX 339--4.}
\vspace{-1.em}
\begin{center}
\begin{tabular}{lp{1.cm}p{1.cm}p{1.cm}p{1.cm}p{1.0cm}}
\hline
\hline

MJD &  Image rms & Measured flux & Flux 3-9  keV & Exposure time & Counts\\
                       &  ($\mu$Jy beam$^{-1}$) &($\mu$Jy beam$^{-1}$) & ($\times 10^{-13}$ erg cm$^{-2}$ s$^{-1}$) & (ks) & \\
\hline
58222		&18.6	&51.2 & --	& -- & --\\   
58369		&40.1	&63.6 & --	&-- &--\\ 
58370       & --    	& --  & 1.29 $\pm$	0.13 & 1.1 & 4	\\
58375		&26.2	&13.2 & --	&-- & --\\      
58382		&26.2	&40.2 &  1.38 $\pm$ 0.14 &2.1 & 10 \\
58389		&35.3	&71.6 & 3.35 $\pm$ 0.33 & 1.9 & 10	\\
58396		&27.7	&43.1 & 3.76 $\pm$ 0.38 & 1.8 & 16	\\               
58402		&27.5	&82.2 & --	& --&--\\ 
58403		&44.8	&47.8 & 3.0 $\pm$ 0.3 &1.8 & 17	\\
58410		&28.2	&17.2	& 0.83 $\pm$ 0.083  & 1.6& 4 \\
58417       & --    & --     	& 1.52 $\pm$ 0.15	& 2.2 &9	\\   
58418		&25.9	&63.4& --	&-- &-- \\                               
58423       & --    & --      	& 8.26 $\pm$ 0.83 & 2.0 & 9		\\
58425		&27.3	&74.2 & -- &-- &--	\\                           
58432		&26.5	&92.1 & -- &-- &--	\\  
58439		&32.3	&76.2 & -- &-- &--	\\  
\hline
\end{tabular}
\end{center}
\label{tab:obs}
\vspace{-3em}

\end{table}%

\subsection{\textit{Swift} X-ray observations}

With the aim of studying the X-ray and radio correlation, we used available data taken by the \textit{Swift} XRT instrument \citep{2000burrows}. For our analysis,  we include measurements made from September 2018 through November 2018 where the source was in the quiescent state. We used eight observations from XRT in photon counting (PC) mode that are close in time to our radio observations. The counts range between 4-17 per observation. Photon pile-up is negligible with such low photon count rates. 

We used the output of the standard pipeline processing and we analyzed the data using the XSPEC software package \citep{1996arnaoud}. We fit the energy spectra with a power-law model accounting for interstellar absorption and we applied Cash statistics to obtain the  X-ray flux from the combined and the individual spectra. We fixed the column density to $N_{H}$ = 6 $\times$ 10$^{21} \rm{cm^{−2}}$ \citep{2004zdziarski,2011cadolle,2013corbel} to obtain reliable fit and to constrain the flux and the photon index $\Gamma$ for the combined spectra. The resulted photon index $\Gamma$ = 2.2 was additionally fixed for the individual spectra fits. All measurement IDs and the obtained fluxes are presented in Table \ref{tab:obs}.

\section{Results}\label{res}
\subsection{Detection in quiescence}

\begin{figure}
\includegraphics[width=0.5\textwidth]{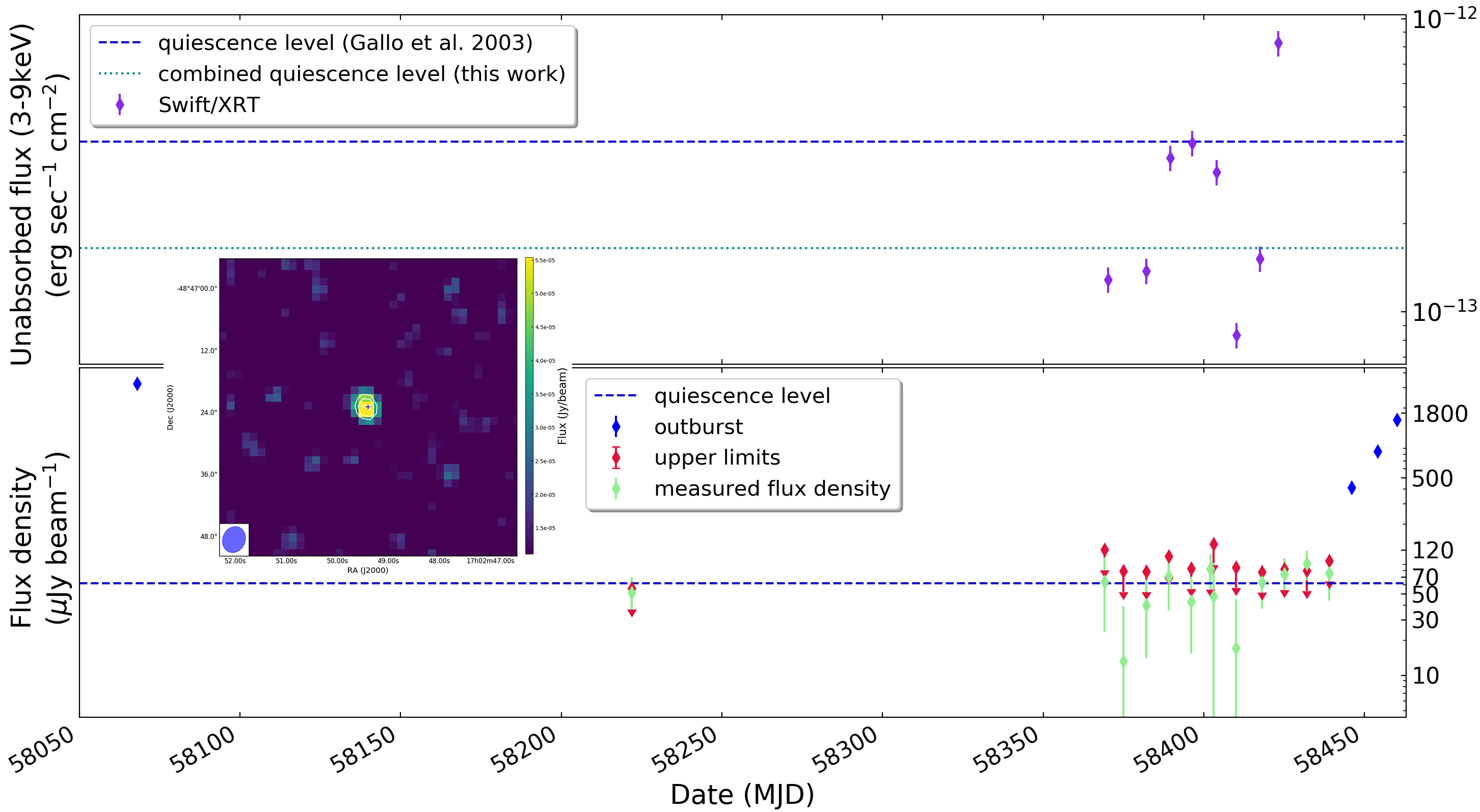}
\vspace{-1.em}
\caption{ The MeerKAT image showing the detection of GX 339--4 in quiescence and  the light curves during the quiescent state. The lower panel shows the radio lightcurve from MeerKAT observations with red arrows indicating the upper limits while the green points correspond to the measured flux during this state. The blue dashed line shows the quiescent level. The blue points show the preceding and following outburst phase \citep[][full outburst will be discussed in a separate study]{2018ATel12287....1T}. In the upper panel, the blue line denotes the low luminosity state as measured by Chandra \citep{2003galloatel} and the cyan line corresponds to the flux as measured by the combined \textit{Swift}/XRT spectra of the observations shown in Table \ref{tab:obs}.  }
\label{fig:quiscpl}
\end{figure}

We started a weekly monitoring of the black hole XRB GX 339--4 at the end of its 2018 outburst.  %
We obtained in total 13 epochs of the GX 339--4 field with MeerKAT (Table \ref{tab:obs}). The source was not detected in any of the individual epochs due to its low flux density. We concatenated the individual epochs in the uv-plane and we obtained an image with an rms background noise level of 11 $\mu$Jy beam$^{-1}$. Owing to the MeerKAT sensitivity and the visibility stacking technique, we obtained a solid detection of the source at  62 $\mu$Jy beam$^{-1}$ during its quiescent state by combining 13 epochs, %(corresponding to the 13 red points in  Fig. \ref{fig:quiscpl}) 
for a total integration time of  $\sim$5 hours (Fig. \ref{fig:quiscpl}). %The source was detected in quiescence state with a peak flux of 62 $\mu$Jy beam$^{-1}$ (Figure \ref{fig:quis} left). 

Figure  \ref{fig:quiscpl}  shows the 3$\sigma$ upper limits in red of the individual MeerKAT observations while in the blue dashed line, we plot the detection level of the stacked image.  In the top panel, we plot the fluxes that were obtained from the quasi - simultaneous observations with the \textit{Swift}/XRT instrument (3-9 keV). 
The source was also detected in the (3-9keV) individual XRT images.  %Figure \ref{fig:quis_swift} shows the combined image of the 8 observations presented in Table \ref{tab:swift}. 
Thanks to the high sensitivity that characterizes the MeerKAT radio telescope and the \textit{Swift} observations, we extend the radio:X-ray correlation for GX 339--4 down to  L$_X$ $\sim$ 10$^{-13}\rm{~erg~cm^{-2}~s^{-1}}$, thus fully covering this system' s evolution cycle.

We have also checked whether the detection in quiescence is dominated by the observations close to the new outburst that started at the end of November 2018 (MJD 58446) (Fig. \ref{fig:quiscpl}). Therefore, in the combined images, we also excluded the the brightest X-ray detection (MJD 58423). The result remained unchanged even without including this observation for the \textit{Swift} image. The same test was performed for the MeerKAT radio map by excluding observing epochs close to the quick rise of the 2018 outburst \citep{2018ATel12287....1T}. In particular, we performed a stacking analysis with MeerKAT data by excluding the last 3 epochs (MJD 58425, 58432 and 58439), where GX 339--4 was detected at the level of $\sim$ 57 $\mu$Jy ($\sim$ 4.2$\sigma$) neglecting the last 3 epochs.

Furthermore, we searched for radio variability hints on $\sim$ 1 month time-scale combining some of the epochs (MJD: 58369-58389, 58396-58418, and 58423-58439), where the signal to noise ratio (S/N) allows 3$\sigma$ detection of the source. The radio variability searches resulted in non-significant change of the flux density above 1 $\sigma$ of the average flux. %We did not detect any significant polarized emission. 
%Although, we notice some short term flux changes in the X-rays, our radio data do not allow possible radio variability to be constrained. %the sensitivity of the radio data is not enough to constrain this.  

%\subsubsection{Radio spectrum}	
The concatenated image was deconvolved over four frequency chunks and hence we were able to obtain a frequency cube that allowed us to measure the flux at each frequency and consequently calculate the spectral index of GX 339--4. 
The four frequency chunks were centered at 962.9 MHz, 1.17 GHz, 1.39 GHz and 1.6 GHz and the flux densities that we obtained were 101.3$\pm$3, 84.9$\pm$5, 75.4$\pm$5, 57.0$\pm$8 $\mu$Jy beam$^{-1}$ , respectively. This results in a negative spectral index $\alpha$ $\sim$ -0.8$\pm$0.4 in contrast to the constantly flat spectra ($\alpha$ $\sim$ 0.35) that we see during the 2018--2019 outburst (Tremou et al. in prep.) and in the past hard states \citep{2000corbel,2003corbel,2013corbel, 2001fender}. Nevertheless, we note here that the sub-band calibration has not been properly evaluated and hence we are aware that our estimates may include a few percent of calibration error.%..  \textbf{(\textit{check the statistics of $\alpha$ in outburst and provide the number})}%resulting from optically thin emission.....

\subsection{Radio:X-ray Correlation}
Simultaneous radio and X-ray observations of black hole XRBs in quiescence have successfully detected the targeted sources only for three low mass XRBs, namely V404 Cyg \citep{2005gallo,2009hynes}, A0620-00 \citep{2006gallo}, and XTE J1118+480 \citep{2014gallo}. They revealed a ratio of L$_{x}$/L$_{Edd}$ $\leq$ 10$^{-8.5}$, while one high-mass XRB, MWC 656 \citep{2017ribo} shows a ratio of L$_{x}$/L$_{Edd}$ $\sim$ 10$^{-9}$. Past deep ATCA \cite[Australia Telescope Compact Array;][]{1992frater} observations revealed only a marginal detection of GX 339--4 at 5 GHz with a flux density of 73 $\pm$ 16 $\mu$Jy) showing a negative spectral index, $\alpha$= -0.6 \citep{2013corbel}. Our current measurements are consistent with the values from \cite{2013corbel}, denoting the true level in quiescence. 
Although the radio:X-ray correlation seems to continue even in the quiescent state \citep{2017plotkin}, some hints of changes in the nature of the X-ray emission have been observed at low luminosities (L$_{x}$/L$_{Edd}$ $\leq$ 10$^{-5}$, \cite{2007gallo,2015plotkin}). Observations at very low accretion rates are therefore key to determine the properties of the faint jets observed in these regimes, and to constrain the still poorly understood physical processes underlying their generation. 

\begin{figure*}
\includegraphics[width=0.913\textwidth]{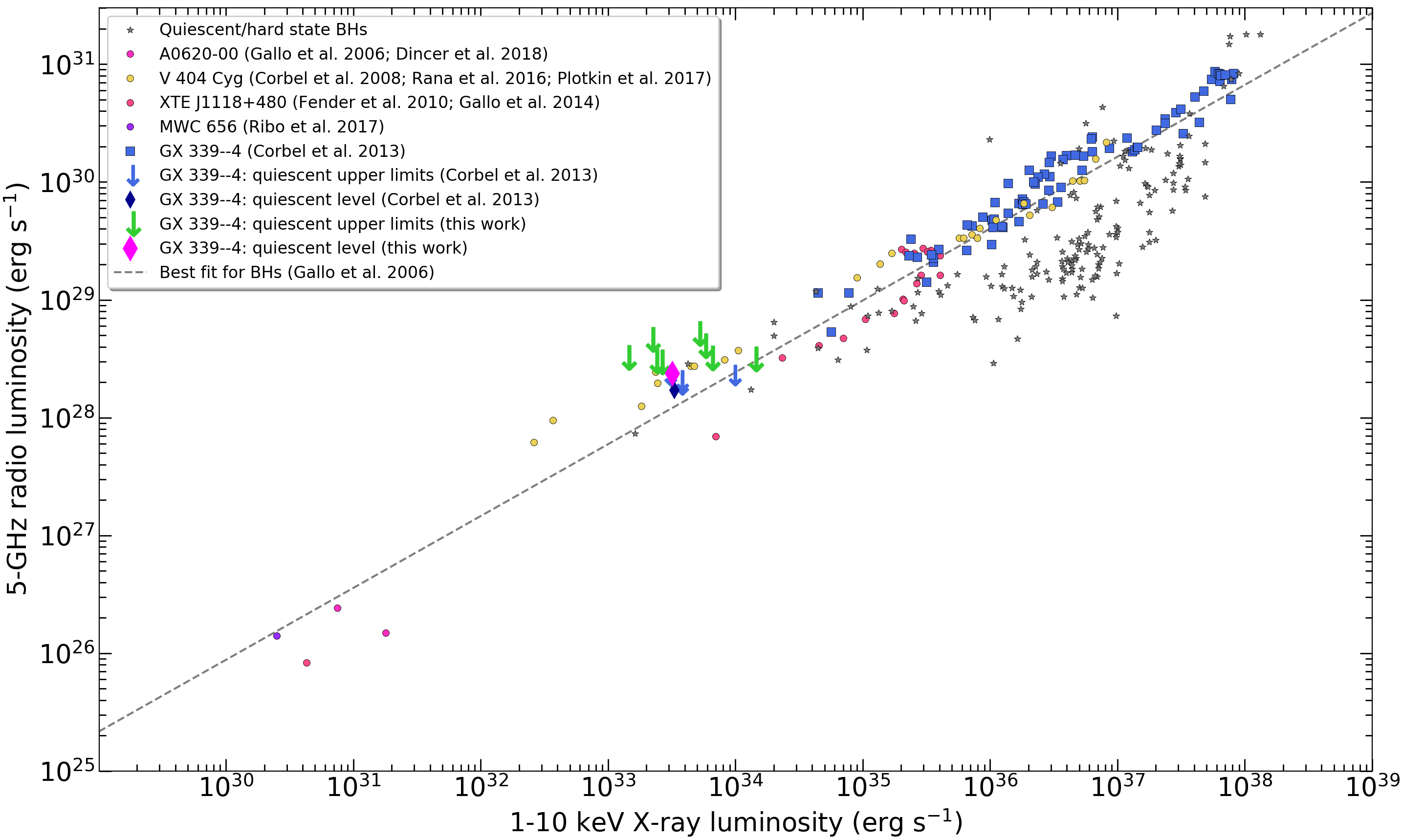}
\vspace{-1.em}
\caption[The L$_{X}$ - L$_{R}$ correlation of quiescent/hard state black holes]{The L$_{X}$ - L$_{R}$ correlation of quiescent/hard state black holes in grey using the database by \cite{2018bahramian}.
GX 339--4 is shown in blue squares (outburst) and blue arrows (upper limits) from \cite{2013corbel}. 
The individual simultaneous measurements of MeerKAT and \textit{Swift}/XRT of this study are shown in green arrows and in purple (diamond), we show the detection in the quiescent state. The blue one denotes the quiescence level from the deep study that has been presented in \cite{2013corbel} 
while the gray dashed line corresponds to their fit with a function of the form L$_{R}$ $\propto$ L$_{X}^{0.61\pm 0.01}$. Quiescent black holes (A0620-00, V404 Cyg, XTE JJ1118+480 and MWC 656), that have been detected, are also over-plotted in circles.}
%\vspace{-1.3em}
\label{fig:lxlr_lc}
\end{figure*}

In Figure \ref{fig:lxlr_lc}, we place our quasi-simultaneous radio and X-ray measurements from this study on the L$_{X}$ - L$_{R}$ plane. In order to compare our results with the population of the quiescent/hard state black holes we we convert the MeerKAT radio flux into radio luminosity at 5GHz, assuming a distance of 8 kpc and a flat spectral index for consistency with objects at hard state. Assuming the same distance, we used the \textit{Swift}/XRT spectra fitting to obtain the unabsorbed X-ray flux (1-10 keV) and we converted it into X-ray luminosity (1-10 keV). % using the %\href{https://heasarc.gsfc.nasa.gov/cgi-bin/Tools/w3pimms/w3pimms.pl}{WebPIMMS tool}. 
Our data points (green and purple) probe the lower part of the correlation while the blue and orange points show the previous extensive study of GX 339--4 during its past outburst-quiescence cycles \citep{2013corbel}. Green arrows show the upper limits obtained by every individual observation reported in this work, while the purple point indicates the level that we obtain from the concatenated image from MeerKAT in radio and \textit{Swift}/XRT in X-rays. 
The gray dashed line represents the best fit for black holes with a function L$_{R}$ $\propto$ L$_{X}^{0.61\pm 0.01}$ \citep{2006gallo}, which is consistent with the fit from the GX 339--4 data presented in \cite{2013corbel} using a higher radio frequency (9 GHz) than the one presented in this work (1.28 GHz). However, in the case of a flat radio spectrum the results do not vary.  

\section{Discussion \& Conclusions}\label{disc}

We have presented an X-ray and a radio analysis of the black hole XRB GX 339--4 in quiescence using observations from \textit{Swift}/XRT and MeerKAT. The source was detected during this state in both X-rays and radio. In order to improve our sensitivity, we concatenated data from several epochs from \textit{Swift}/XRT and from MeerKAT, and we detected the source at the level of 1.6 $\times$10$^{13}\rm{~erg~cm^{-2}~s^{-1}}$ and 62 $\mu$Jy beam$^{-1}$, respectively. 

Sampling the low-luminosity end of the radio:X-ray correlation, we probe low Eddington accretion rates of XRBs at the low X-ray luminosity quiescent level of the order of 10$^{33}\rm{~erg~s^{-1}}$. The radio:X-ray correlation of L$_{X}$ $\propto$ L$_{R}^{0.62\pm0.01}$ in GX 339--4 has been well constrained for the brightest hard states by \cite{2013corbel} using measurements covering $\sim$ 15 years. Our measurements confirm that the same correlation seems to continue with no break down to low luminosities, where we detect GX 339--4, which is characterized by a soft X-ray spectrum. 
%Corbel et al 2013 and references therein \cite{2006corbel}  --> upper end of the quiescence state. 
%Possible scenarios that can explain the emission mechanism during the quiescence state comprises of synchrotron cooled jet that dominates the emission or Synchrotron Self-Compton (SSC) due to the hot accretion flow or from a radiatively cool jet and/or the inefficient particle acceleration along the jet axis. 
While the binary system is going towards quiescence from the hard state, the X-ray spectral shape becomes softer until it reaches a constant shape \citep[$\Gamma$ $\sim$ 2.1,][]{2006corbel,2008corbel,2013plotkin,2017plotkin}. Our fitting of the combined \textit{Swift}/XRT spectrum constrains the photon index $\Gamma$ value which is consistent with the soft X-ray spectra.  The soft X-ray spectra favor a mechanism where in the inner regions of the accretion disk, radiatively inefficient outflows are likely to develop within a geometrically thick and hot area. %, where compact jets are believed to be formed.
%GX339--4 shows a robust disk- jet coupling since the slope between different phases is maintained at the same level. 

The slope of the radio:X-ray correlation 
for GX 339--4 is similar to that of the well studied binary source V404 Cyg, which has not shown any evidence that a synchrotron cooled jet could dominate the X-ray emission \citep{2017plotkin}. On the other hand, synchrotron self-Compton (SSC) processes from a radiatively cooled jet or a hot accretion flow may be responsible for generating the X-ray emission \citep{2014poutanen,2016malzac}. Furthermore, the X-ray emission in quiescence can be driven by a less efficient particle acceleration along the jet axis implying optically thin synchrotron emission by non-thermal particles \citep{2013plotkin,2017connors}.

MeerKAT radio observations show a negative spectral index $\alpha$ = -0.8$\pm$0.4 in the quiescent state favoring an optically thin emission in contrast to the constantly flat spectra ($\alpha$ $\sim$ 0.35) that seems to dominate the outburst phase of GX 339--4 (Tremou et al. in prep.). Although, radio emission in quiescence denotes the presence of hard state jets, the wide range of spectral indices that have been previously seen, it is not fully understood \citep{2019plotkin} and it may be related with non-canonical jet geometry. Slightly negative spectral index could be seen in the case of a decelerating or a slowly expanding jet, while an inverted spectrum \citep[eg: A0620-00;][]{2018dincer} could be generated due to the fast expanding parts of the jet (outer regions of the jet). 

During our ThunderKAT campaign over the next years, we will be able to constrain the spectral evolution during both the outburst and quiescent state, the latter of which has been poorly understood so far. 

\section*{Acknowledgements}

ET and SC acknowledge financial support from the UnivEarthS Labex program of Sorbonne Paris Cit\'{e} (ANR-10-LABX-0023 and ANR-11-IDEX-0005-02). JCAM-J is the recipient of an Australian Research Council Future Fellowship (FT140101082), funded by the Australian Government. PJG acknowledges support from the NRF SARChI program under grant number 111692. PAW acknowledges support from UCT and the NRF.
We acknowledge the use of data obtained from the High Energy Astrophysics Science Archive Research Center (HEASARC), provided by NASA' s Goddard Space Flight Center. We thank the staff at the South African Radio Astronomy Observatory (SARAO) for scheduling these observations. The MeerKAT telescope is operated by the South African Radio Astronomy
Observatory, which is a facility of the National Research Foundation,
an agency of the Department of Science and Innovation. 
This work was carried out in part using facilities and data processing pipelines developed at the Inter-University Institute for Data Intensive Astronomy (IDIA). IDIA is a partnership of the Universities of Cape Town, of the Western Cape and of Pretoria. %We acknowledge the use of the Nan\c cay Data Center computing facility  (CDN - Centre de Donn\'{e} es de Nan\c cay). The NDC is hosted by the Station de Radioastronomie de Nan\c cay in partnership with Observatoire de Paris, Universit\'{e} d' Orl\'{e}ans, OSUC and the CNRS. The CDN is supported by the Region Centre Val de Loire, d\'{e}partement du Cher.
We acknowledge the use of the Nan\c cay Data Center, hosted by the Nan\c cay Radio Observatory (Observatoire de Paris-PSL, CNRS, Universit\'{e} d' Orl\'{e}ans), and also supported by Region Centre-Val de Loire.

%%%%%%%%%%%%%%%%%%%%%%%%%%%%%%%%%%%%%%%%%%%%%%%%%%



\bibliographystyle{mnras}
\nocite{2010fender,2016rana,2018dincer}
\bibliography{gx339_4-q.bib} % if your bibtex file is called example.bib

%\begin{thebibliography}{99}
% \bibitem[\protect\citeauthoryear{Author}{2012}]{Author2012}
% Author A.~N., 2013, Journal of Improbable Astronomy, 1, 1
% \bibitem[\protect\citeauthoryear{Others}{2013}]{Others2013}
% Others S., 2012, Journal of Interesting Stuff, 17, 198
%\end{thebibliography}

%%%%%%%%%%%%%%%%%%%%%%%%%%%%%%%%%%%%%%%%%%%%%%%%%%

%%%%%%%%%%%%%%%%% APPENDICES %%%%%%%%%%%%%%%%%%%%%

%\appendix



%%%%%%%%%%%%%%%%%%%%%%%%%%%%%%%%%%%%%%%%%%%%%%%%%%

% Don't change these lines
\bsp	% typesetting comment
\label{lastpage}
\end{document}